\documentclass[a4paper,11pt]{article}
\usepackage{pos}
\usepackage{wrapfig}
\usepackage{graphicx}
\newcommand{\nss}{$\nu$SpaceSim}
\newcommand{\nut}{$\nu_\tau$ }
\usepackage{placeins}
\usepackage{caption}
\usepackage[numbers,sort&compress]{natbib}
\usepackage{hyperref}
\usepackage{orcidlink}
\usepackage{subcaption} 
\newcommand{\Offline}{$\overline{\textrm{Off}}$\hspace{.05em}\protect\raisebox{.4ex}{$\protect\underline{\textrm{line}}$}}

\title{$\nu$SpaceSim: An end-to-end simulation package to model the sensitivity of UHECR experiments to upward-moving extensive air showers sourced by cosmic neutrinos interacting in the Earth}

\ShortTitle{$\nu$SpaceSim}

\author*[a]{Jorge Caraça-Valente Barrera \orcidlink{0009-0000-1677-3639}}
\onbehalf{on behalf of the $\nu$SpaceSim Collaboration}

\affiliation[a]{Colorado School of Mines, Department of Phisics\\
  1500 Illinois St, Golden, CO, USA}

\emailAdd{jcaracavalente@mines.edu}

\abstract{Neutrinos act as probes of hadronic processes and offer a distinctive view into their astrophysical origins at high energies. When reaching energies on the PeV scale, $\nu_\tau$ interactions within the Earth can produce a significant flux of $\tau$-leptons. These $\tau$-leptons subsequently decay, generating upward-moving extensive air showers (EAS). Using the Earth as a target for neutrinos and the atmosphere as a signal generator effectively creates a detector with a mass $\gg$ gigaton. $\nu$SpaceSim is a comprehensive simulation developed to model all the relevant physical processes that describe the neutrino-induced, Earth-emergent lepton chain. The simulation models neutrino interactions inside the Earth that produce leptons, the propagation of the leptons through the Earth into the atmosphere, and their decay, forming composite EAS. Next, it models the generation of air optical Cherenkov and radio signals from these showers, including the propagation and attenuation of these signals through the atmosphere, accounting for effects such as clouds and the ionosphere. Finally, the simulation models the detector response according to the parameters defined by the user (such as altitude, effective area, frequency band...). Through this end-to-end simulation, $\nu$SpaceSim aims to help design the next generation of balloon- and space-based experiments, to estimate the exposure of ground-based experiments to these showers, and to understand the data from recent experiments such as EUSO-SPB2 and ANITA. 
}

\FullConference{7th International Symposium on Ultra High Energy Cosmic Rays (UHECR2024)\\
 17-21 November 2024\\
Malargüe, Mendoza, Argentina\\}

\begin{document}
\maketitle

\section{Introduction}
The high energy astrophysical phenomena (VHE: $E_\nu \gtrsim 1 $\,PeV) can be observed mainly through four known distinct messengers: cosmic rays (made up of nuclei such as protons or iron), gamma rays, gravitational waves, and neutrinos. Combining these four messengers may help us find and understand the mechanisms behind the astrophysical sources of the highest energies \cite{pathways2021}. In particular, due to the neutral and weakly interacting nature of neutrinos, their observation may grant us the opportunity to detect sources beyond regions that are opaque to electromagnetic radiation (gamma rays) or that present a magnetic field that deflects all charged particles (making it very difficult to retrace the cosmic ray back to its source) \cite{magfields,whitepaper}. On top of that, there are many more sources of $\nu$ than detectable gravitational waves.

\begin{wrapfigure}[15]{r}{0.5\textwidth} 
    \vspace{-2mm}
    \centering
    \includegraphics[width=0.5\textwidth]{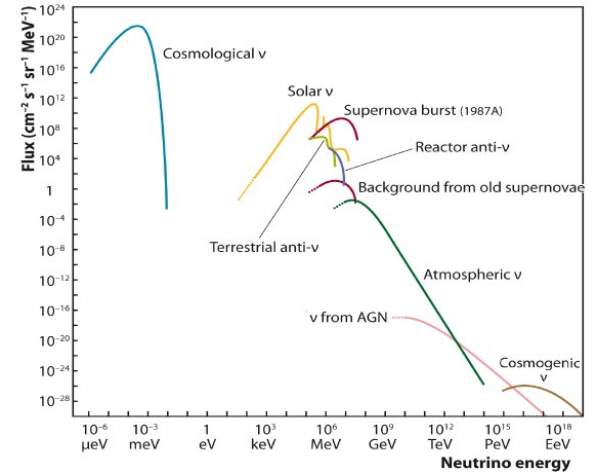} 
    \vspace{-7mm}
    \caption{Measured and expected fluxes of the main neutrino sources as a function of energy. \cite{sources}}
    \label{fig:sources} 
\end{wrapfigure}
\indent Figure \ref{fig:sources} shows the predicted flux of neutrinos as a function of energy and their sources. The VHE region includes neutrinos from Active Galactic Nuclei (AGN) and cosmogenic neutrinos. However, the flux of these neutrinos is still highly suppressed. When also considering that neutrinos have an extremely low cross-section with matter \cite{crosssections}, the direct detection of VHE neutrinos becomes practically unfeasible. Therefore, an indirect approach is needed, as it enables a radically larger target detection volume to be used. The IceCube Collaboration \cite{icecube} has done this by using $\sim\,1$\,km$^3$ of antarctic ice, and has obtained several significant results. IceCube provided evidence for an extra-solar astrophysical neutrino flux, with energies extending from $E_\nu>10$\,TeV to possibly $\sim$10\,PeV \cite{ic1}, including an event matching the Glashow resonance at 6.3\,PeV \cite{ic2}. They also detected a correlation of neutrinos with the nearby active galaxy NGC 1068 \cite{ic3}, the blazar flare TXS 0506+056 \cite{ic4}, and with Tidal Disruption Events \cite{ic5}. This demonstrates the existence of extra-galactic neutrino sources and highlights the importance of multi-messenger astrophysics. It also proves the success of the indirect detection method for VHE neutrinos, and its necessity to observe and better understand transient astrophysical events. 

A complementary approach to IceCube that these findings motivate is the development of sub-orbital balloon experiments, such as EUSO-SPB2 \cite{spb2}, PUEO \cite{pueo} or PBR \cite{pbr}, and space-based experiments, such as POEMMA \cite{poemma}. These experiments are designed to use the entire mass of the Earth as the target volume for neutrino interactions, and the atmosphere as the detection volume, drastically increasing the sensitivity to Ultra High Energy (UHE: $E_\nu \gtrsim 1$\,EeV) neutrinos. 

At the sources the initial ratio of neutrino flavour is $N_{\nu_e} : N_{\nu_\mu} : N_{\nu_\tau} \sim 1 : 2 : 0$ \cite{flavours}. However, neutrino oscillations over astronomical distances causes the neutrino flavour ratio that arrives at Earth to balance out to be $1:1:1$ \cite{oscillations}. To detect $\nu$-induced upward-going EAS, first a $\nu_\tau$ must reach the Earth and cross through it, undergoing Charged Current (CC) interactions that produce a $\tau$ lepton or Neutral Current (NC) interactions that produce a lower energy $\nu_\tau$. The $\tau$ lepton then decays; if this happens inside the Earth it produces another $\nu_\tau$ which may have another CC interaction to produce a new $\tau$ (regeneration effect). If the $\tau$ makes it out of the Earth, it can decay in the atmosphere and its products will then interact with the atoms in the atmosphere to initiate an EAS. This EAS will then produce optical (Cherenkov and fluorescence) and radio signals that propagate through the atmosphere until they reach the detector, be it on the ground, onboard a balloon, or in space. The $\nu_\tau$ is the main candidate for this detection channel due to the short half-life of the $\tau$, ensuring it decays in the atmosphere. Conversely, for $\nu_e$ the $e^\pm$ interaction length is too short for a significant fraction of them to make it out of the Earth, while for $\nu_\mu$ both the shorter interaction length of the $\mu$ and the longer lifetime in the atmosphere significantly suppress its detection. However, the muonic channel can still be considered, albeit only one $\nu_\mu \rightarrow \mu$ interaction is required since muons quickly lose energy in transit through the Earth. Though 
\begin{wrapfigure}[23]{r}{0.4\textwidth} 
    \centering
    \includegraphics[width=0.38\textwidth]{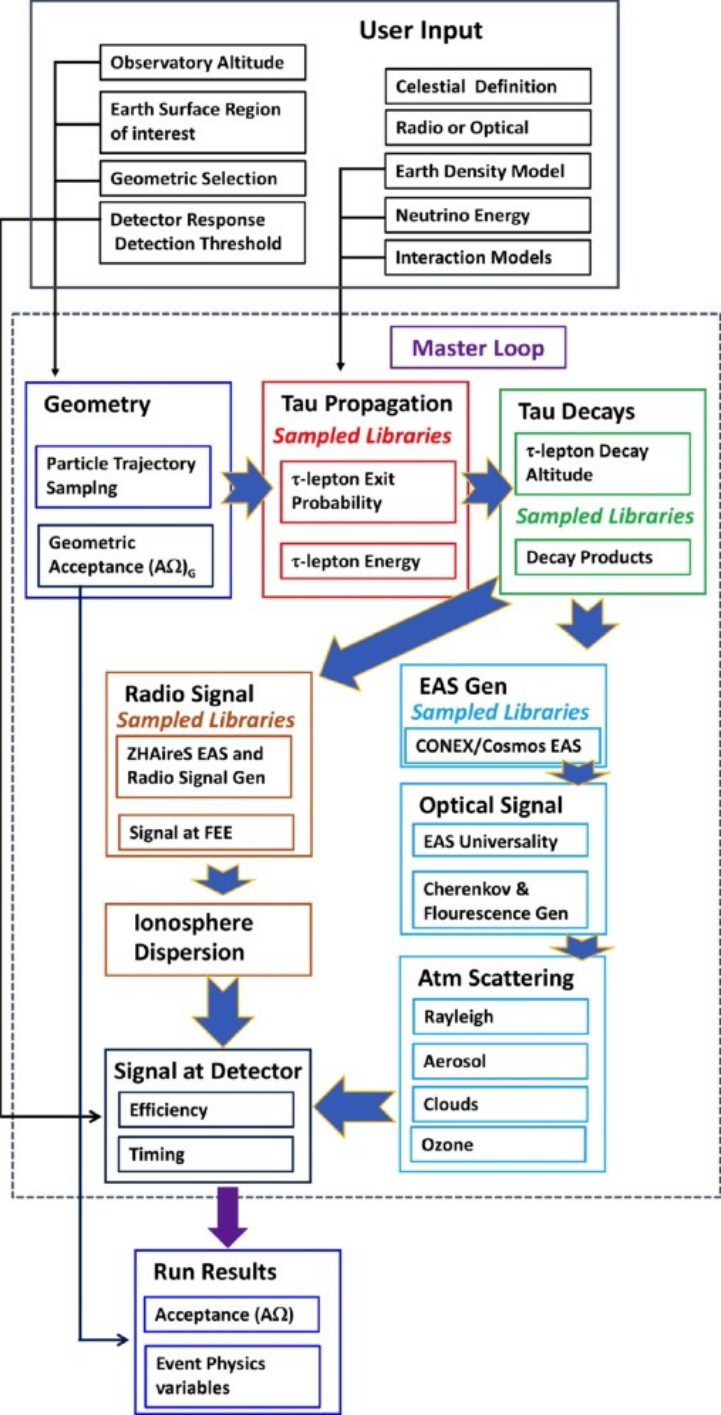} 
    \vspace{-2mm}
    \caption{$\nu$SpaceSim flowchart, showing the modularity of the simulation package}
    \label{fig:chain} 
\end{wrapfigure}\indent The need to accurately model the sensitivity of next-generation cosmic neutrino experiments, including their response to neutrino-induced signals, motivates the creation and continued development of \nss. This comprehensive, end-to-end simulation software package is designed to efficiently provide detailed modeling of neutrino interactions, lepton propagation, and detector responses, enabling precise predictions and optimizations for upcoming experiments. 

\section{Simulation framework}
\begin{figure}[h!]
    \centering
    \includegraphics[width=0.94\textwidth]{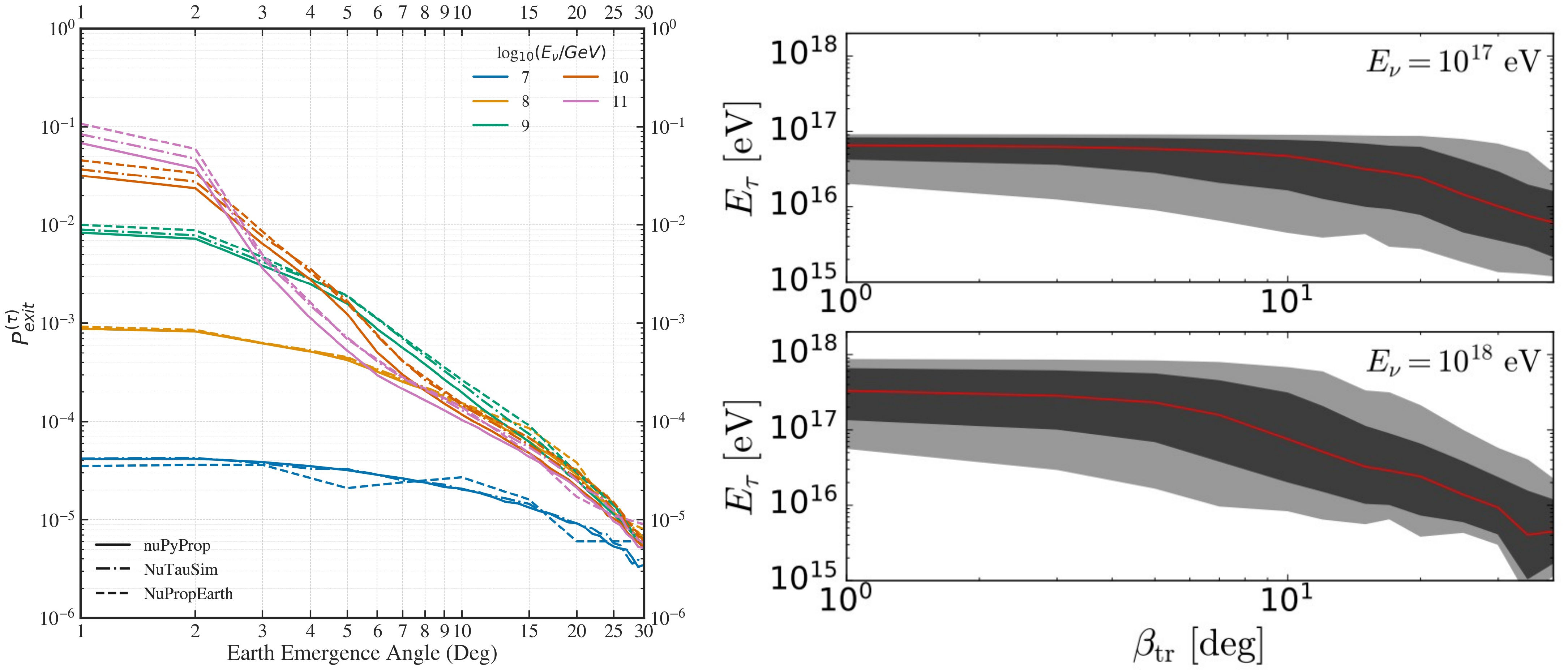} 
    \vspace{-1mm}
    \caption{Left: Earth exit probability of the $\tau$ for different \nut energies and Earth emergence angles. \cite{exitprob}. Right: Average energy of the outgoing $\tau$ and range of energies including 68\% (black) and 95\% (grey) of the events as a function of Earth emergence angle, for a $E_{\nu_\tau}=0.1$\,EeV on the top and 1\,EeV on the bottom \cite{exitenergy}.  }
    \label{fig:exit} 
\end{figure}

\begin{figure}[h]
    \centering
    \includegraphics[width=1\textwidth]{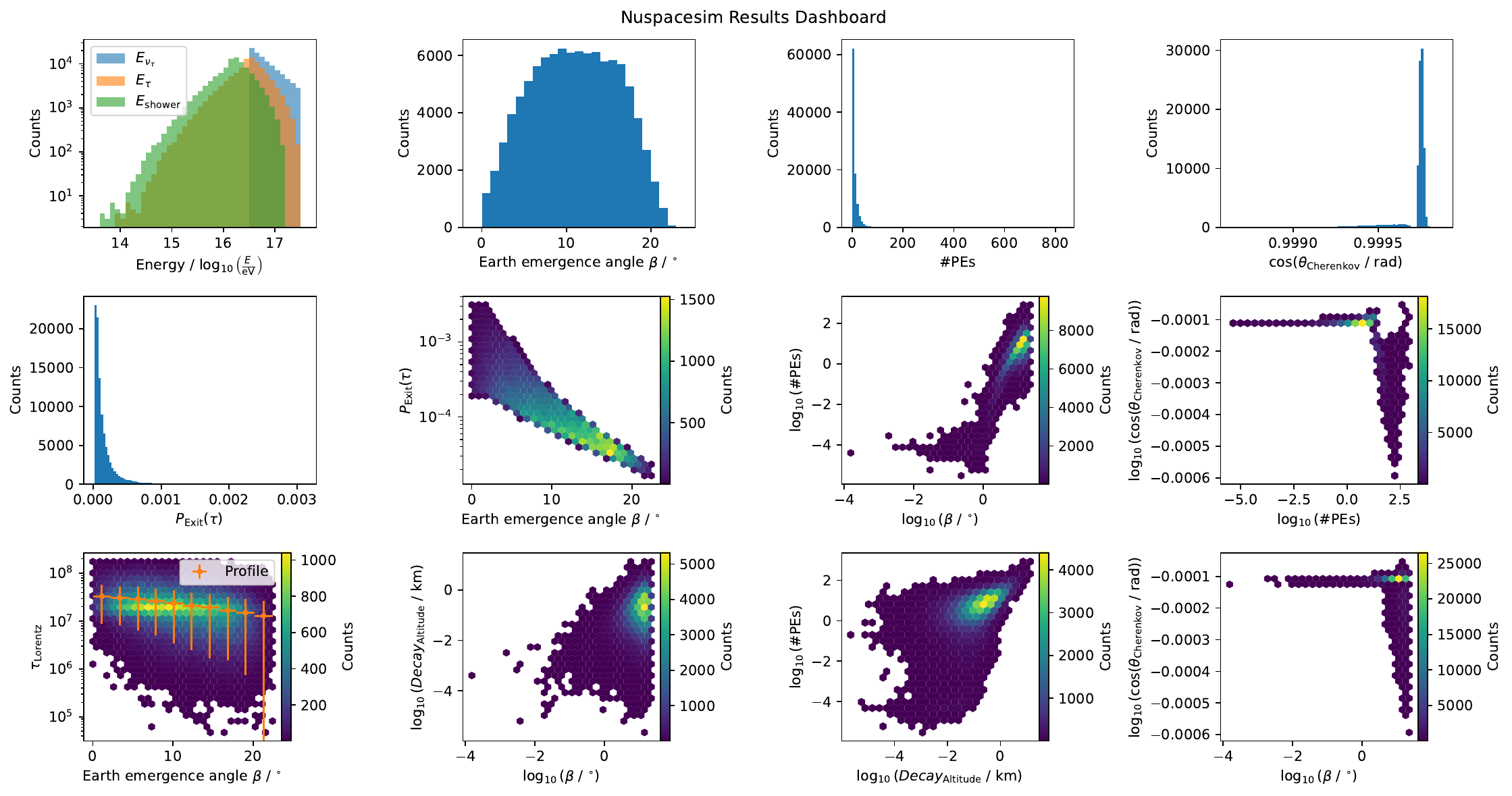} 
    \caption{Dashboard exemplifying some of the plots produced by $\nu$SpaceSim }
    \vspace{-2mm}
    \label{fig:plots} 
\end{figure}

$\nu$SpaceSim uses a vectorized Python wrapper with different modules also written in Python. Figure \ref{fig:chain} shows the flowchart of $\nu$SpaceSim and the modules that are part of it. For a user-given detector configuration, it generates the geometry of the incoming neutrinos. It models the propagation of the $\nu_\tau$ and $\tau$ inside the Earth, including the CC and NC interactions that may occur. It calculates the probability of the $\tau$ exiting the Earth (Figure \ref{fig:exit}), as well as the energy and decay of the resulting $\tau$. The simulation then models the formation of the EAS, the generation of the subsequent optical and radio signals, and their propagation through the atmosphere. This includes accounting for ionospheric dispersion and atmospheric scattering effects, such as ozone, clouds, aerosols, and Rayleigh scattering. Finally, it simulates the arrival of the signal at the detector. The output provides data to calculate the cosmic neutrino sensitivity of the defined instrument and saves event-by-event analysis variables in astropy tables \cite{astropy} that can be plotted (Figure \ref{fig:plots}).

The original modules, written in Fortran, C, and C++, have been 'pythonized' to enhance performance and usability without compromising accuracy. It uses a sampled library methodology to build on the past work of other packages and increase efficiency. This approach relies on pre-built libraries in the shape of lookup tables for the most computationally demanding parts of the simulation. Some of the libraries used are NuPyProp \cite{nupyprop} and NuTauSim\textbackslash NuLeptonSim \cite{nuleptonsim} for the \nut propagation through the Earth, Pythia8 \cite{pythia} for the decay products of the $\tau$, MERRA-2 \cite{merra} for the atmospheric data and ZHAireS \cite{zhaires} for the radio signal modeling. This enables the user to further customize their simulation by varying parameters at will, or even use their own libraries or lookup tables as long as the appropriate input format is followed. It also counts with inherent multi-core processing via Dask \cite{dask}. This approach makes $\nu$SpaceSim extremely fast, generating $10^6$ events in $\sim$\,5\,minutes using a MacBook Pro with an M2Max processor. A precompiled package of version 1.6.0 can be downloaded publicly through pip \cite{pip} or from the HEASARC web portal \cite{heasarc}.
\vspace{-3mm}
 The open-source code is accessible via a public GitHub repository \cite{github}, which provides documentation and a platform for users to report issues.  Source-code tarballs for previous releases are stored and maintained by the HEASARC at GSFC. 

\section{Simulation features and upgrades in progress}
\begin{wrapfigure}[18]{r}{0.35\textwidth}
    \centering
    \vspace{-25mm}
    \includegraphics[width=0.34\textwidth]{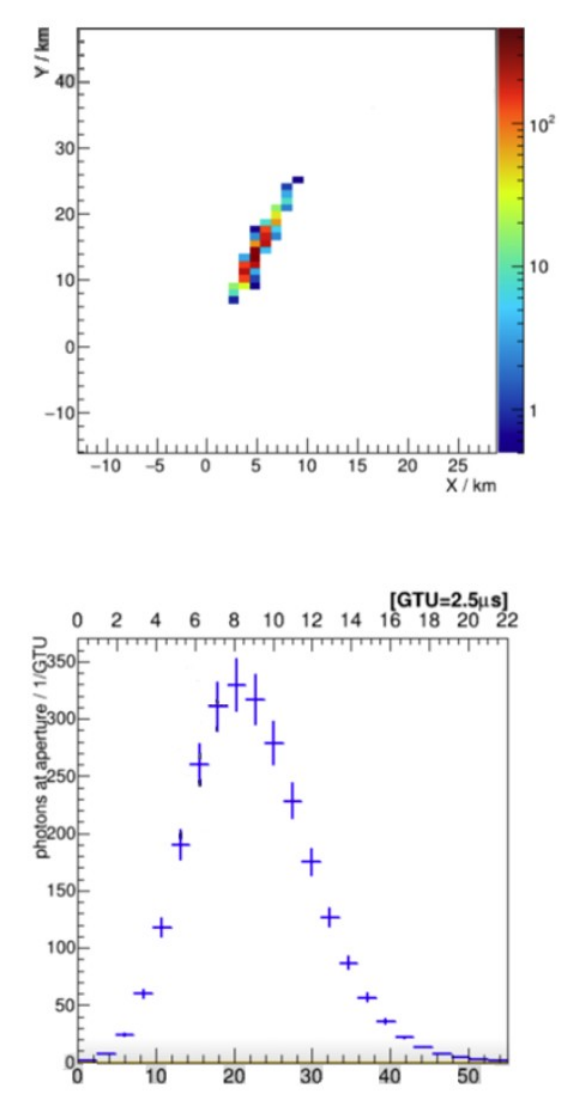} 
    \caption{Upper: triggering pixels in the detection plane. Lower: Shower profile reconstruction.}
    \label{fig:poemma} 
\end{wrapfigure}
The user input format is in TOML to provide a simple and clear way to specify the simulation parameters, and the output is given as either HDF5 or FITS format. There is also the option of outputting the EAS longitudinal profiles of the generated events in a CONEX-like Root format \cite{conex} (currently in beta version). This allows for the direct interfacing of $\nu$SpaceSim output with the simulation framework of experiments from other collaborations (e.g., \Offline\ \cite{offline}) to include the instrument-specific electronics, cameras, etc. \Offline\ was developed and is maintained by the Pierre Auger Collaboration for their Observatory. It was also adapted for the EUSO Collaboration to support balloon-borne or space-based experiments like EUSO-SPB2, PBR, or POEMMA. Figure \ref{fig:poemma} shows an example of a simulation in Euso \Offline\ \cite{eusooffline} of a triggering shower of 100\,EeV generated by $\nu$SpaceSim in a POEMMA-like fluorescence camera at 525\,km of altitude. Current work is in progress in collaboration with the Pierre Auger Observatory to calculate its Fluorescence Detector's exposure to neutrino-induced upward-going EAS. 

\begin{wrapfigure}[14]{r}{0.52\textwidth}
    \vspace{-3mm}
    \centering
    \includegraphics[width=0.51\textwidth]{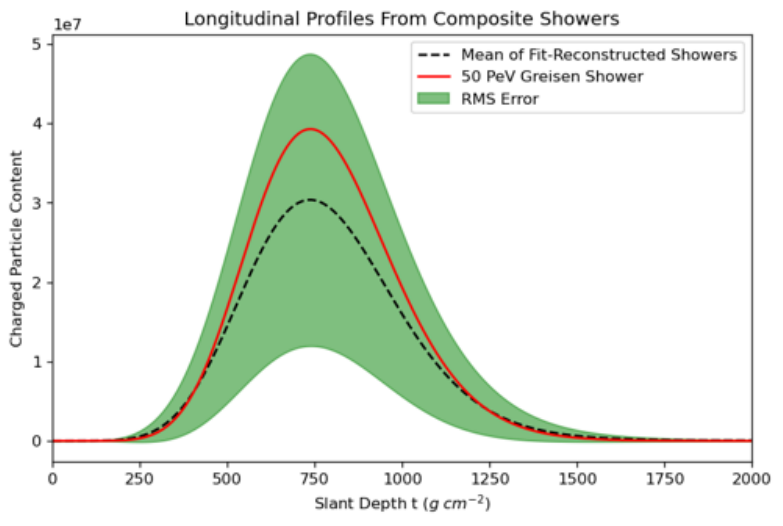} 
    \vspace{-2mm}
    \captionof{figure}{Mean and variance of a composite EAS using Pythia8 and CONEX, along with the Greisen parametrization currently employed in \nss}
    \label{fig:composite}
\end{wrapfigure} 

The default package used to model $\tau$ propagation and exit probability is nuPyProp. However, a version of $\nu$SpaceSim using nuLeptonSim is being developed, and the modularity of the software allows for the use of other publicly available codes such as TauRunner \cite{taurunner} or nuPropEarth \cite{nupropearth}, allowing a user to quantitatively measure the differences and systematic errors in this part of the simulation. $\nu$SpaceSim also models both cosmic diffuse neutrinos and transient neutrino sources while providing the pointing specifications needed for target-of-opportunity follow-up observations.

The EAS module is being upgraded in multiple ways. The modeling of the $\tau$ decay products using Pythia8 is being implemented along with the generation of composite EAS according to said decay products. This will use CONEX-parametrized component showers and will account for variability both in $X_{\textrm{max}}$ and $N_{\textrm{max}}$. Figure \ref{fig:composite} shows an example of this implementation, currently in beta version, displaying the mean, RMS variation, and a Greisen parametrization for reference at the respective energy. Future additions will include optical and radio signals from muonic EAS, signals from the muon component of regular EAS, and different background effects.   
\begin{wrapfigure}[17]{r}{0.52\textwidth}
    \vspace{-3mm}
    \centering
    \includegraphics[width=0.51\textwidth]{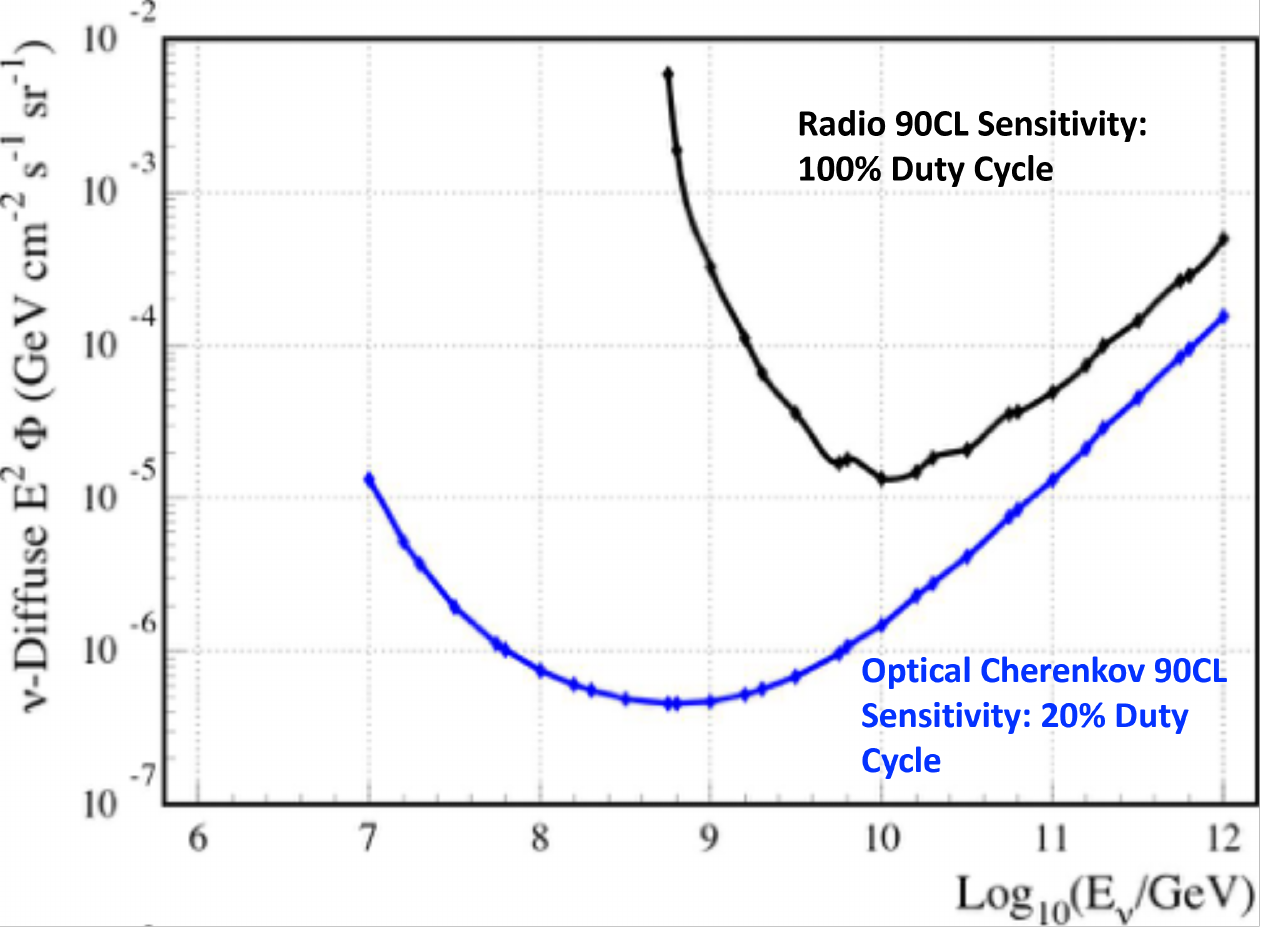}
    \vspace{-2mm}
    \caption{Simulated sensitivity of an SPB-2
like balloon experiment, showing the 90\% CL for both radio and optical Cherenkov.}
    \label{fig:sens}
\end{wrapfigure}\indent The effects of clouds were recently implemented for version 1.5.1. It assumes the clouds are opaque to Cherenkov light and offers the flexibility to use data from the MERRA-2 database, include a constant layer of clouds at a specified altitude, or exclude clouds entirely. The Cherenkov, radio, and timing calculations are also being updated to increase their accuracy and efficiency. In particular, the Cherenkov light will be modeled using CHASM \cite{chasm}, which exploits the universality of charged particles to calculate the Cherenkov photon yield at the given detector position and its angular distribution, by sampling throughout the shower stages and altitudes. The radio signal currently uses ZHAireS.
\begin{wrapfigure}[12]{r}{0.52\textwidth}
    \vspace{-32mm}
    \centering
    \includegraphics[width=0.51\textwidth]{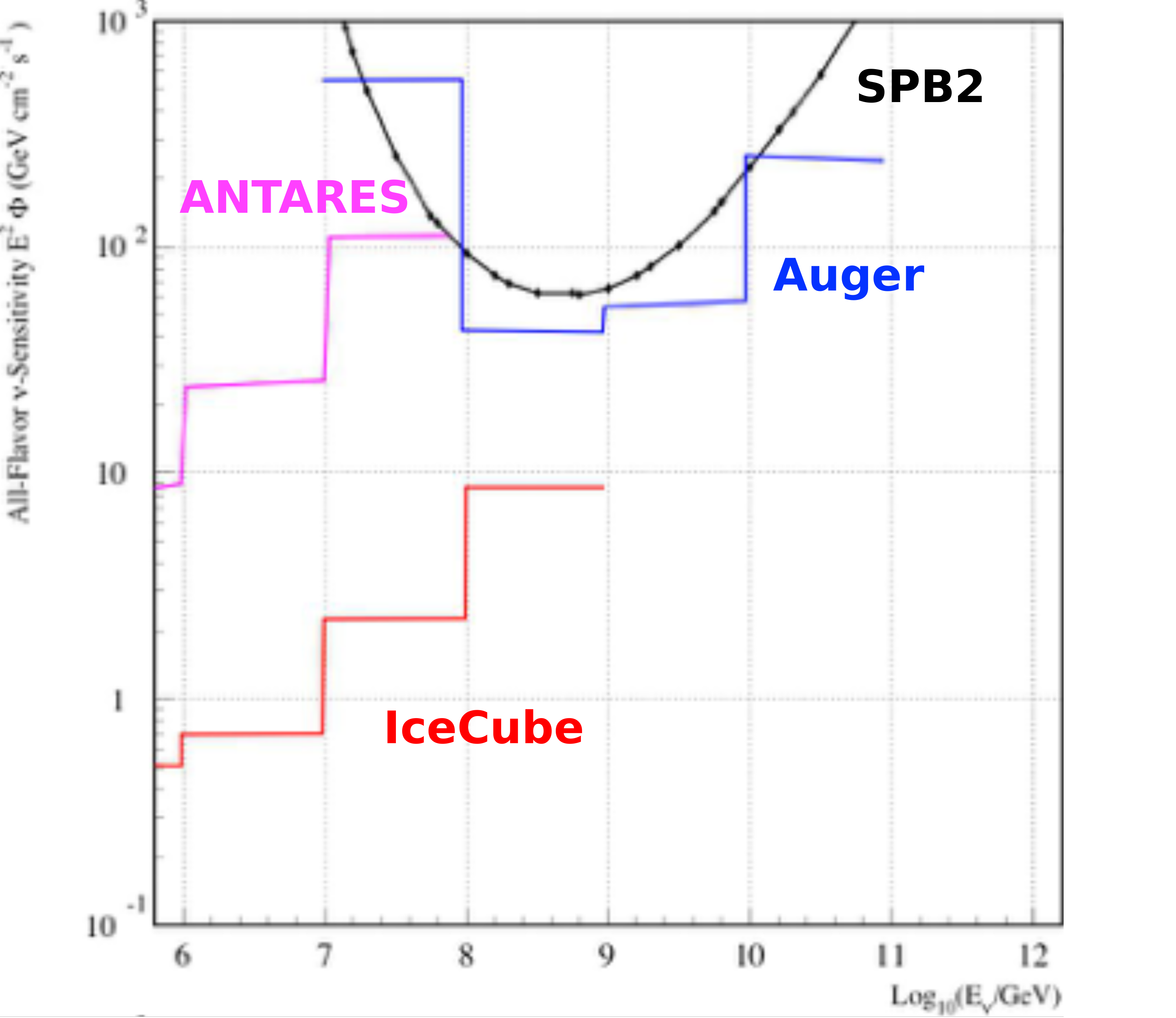}
    \vspace{-2mm}
    \caption{Measured 14-day all-flavor 90\% confidence sensitivity of the ANTARES, IceCube, and Auger experiments to GW170817, and the simulated sensitivity for SPB-2 \cite{search}.}
    \label{fig:gw}
\end{wrapfigure}



\section{$\nu$SpaceSim  results}
$\nu$SpaceSim can be used to calculate the sensitivity of any optical and/or radio instrument in a straightforward and user-friendly way, particularly balloon-borne and space-based experiments. Figure \ref{fig:sens} shows the simulated sensitivity of a balloon mission at 33\,km altitude, pointing between the Earth's limb and 6.4$^\circ$ below, full azimuthal aperture, 1\,m$^2$ optical camera area and 10 radio antennas in the 30-300\,MHz band using $\nu$SpaceSim. All of these parameters and more are customizable, including some of the sampled libraries involved in the modeling of the different physical processes that are part of the simulation.

Using $\nu$SpaceSim's transient source modeling enables a direct performance comparison between balloon missions, such as EUSO SPB-2, and ground-based experiments in observing transient target-of-opportunity events. Figure \ref{fig:gw} shows a simulation of what the 14-day sensitivity of EUSO SPB-2 would have been to the GW170817 binary merger \cite{gw17} compared to the actual sensitivity of other experiments if it had been active during the merger. It shows how this detection method can compare and compete with the biggest observatories available, and demonstrates the importance of having software adapted to a target-of-opportunity program for the detection of transient events.
\section{Conclusion}
$\nu$SpaceSim provides an end-to-end earth-skimming cosmic neutrino simulation that models all the physical processes that lead to the detection of both the optical and radio signals produced by the $\nu_\tau$-induced EAS. The existing features and the upcoming upgrades presented in this paper are continually being developed within $\nu$SpaceSim to provide the VHE and UHE neutrino community with new improvements in modeling accuracy, customization, efficiency, and accessibility. Comments, bug reports, or new feature proposals are always welcome on the GitHub page \cite{github}.

\acknowledgments
 This work is supported by NASA RTOP21-APRA21-0071 at
 NASA/GSFC and NASA grants 80NSSC22K1520 at the University of Chicago, 80NSSC22K1517
 at the Colorado School of Mines, 80NSSC22K1523 at the University of Iowa, 80NSSC22K1519 at
 Pennsylvania State University, and 80NSSC22K1522 at the University of Utah.
{\small
\setlength{\baselineskip}{0.91\baselineskip} 

}
\vspace{-5mm}
\section*{$\nu$SpaceSim Collaboration}
\small

John Krizmanic$^1$, Yosui Akaike$^2$, Luis Anchordoqui$^3$, Douglas Bergman$^4$, Isaac Buckland$^4$, Jorge Caraça-Valente$^5$, Austin Cummings$^6$, Johannes Eser$^7$, Fred Angelo Batan Garcia$^8$, Diksha Garg$^9$, Claire Guépin$^{10}$, Tobias Heibges$^5$, Luke Kupari$^9$, Andrew Ludwig$^{11}$, Simon Mackovjak$^{12}$, Eric Mayotte$^5$, Sonja Mayotte$^5$, Angela Olinto$^7$, Thomas Paul$^{3}$, Alex Reustle$^1$, Andrew Romero-Wolf$^{11}$, Mary Hall Reno$^9$, Fred Sarazin$^5$, Tonia Venters$^1$, Lawrence Wiencke$^5$, Stephanie Wissel$^6$
\vspace{4mm}

\noindent$^1$NASA/Goddard Space Flight Center, Greenbelt, Maryland 20771 USA	$^2$Waseda Institute for Science and Engineering, Waseda University, Shinjuku, Tokyo, Japan	 $^3$Department of Physics and Astronomy, Lehman College, City University of New York, New York, New York, 10468 USA	$^4$Department of Physics and Astronomy, University of Utah, Salt Lake City, Utah 84112 USA	$^5$Department of Physics, Colorado School of Mines, Golden, Colorado 80401 USA	$^6$Department of Physics, Pennsylvania State University, State College, Pennsylvania 16801 USA	$^7$Department of Astronomy and Astrophysics University of Chicago, Chicago, Illinois 60637 USA	$^8$Department of Physics, Columbia University, New York, New York USA	$^9$Department of Physics and Astronomy, University of Iowa, Iowa City, Iowa 52242 USA	$^{10}$Laboratoire Univers et Particules de Montpellier (LUPM) France	$^{11}$Jet Propulsion Laboratory, California Institute of Technology, Pasadena, California 91109, USA	$^{12}$Institute of Experimental Physics, Slovak Academy of Sciences, Kosice, Slovakia


\begin{thebibliography}{99}
\setlength{\itemsep}{0em} 



\bibitem{pathways2021}
National Academies of Sciences, Engineering, and Medicine. (2021). \textit{Pathways to Discovery in Astronomy and Astrophysics for the 2020s.}

\bibitem{magfields}
Brandenburg, A., \& Subramanian, K. (2005). Astrophysical magnetic fields and nonlinear dynamo theory. Physics Reports, 417(1-4), 1-209.
\bibitem{whitepaper}
Ackermann, M. et al. (2022). High-energy and ultra-high-energy neutrinos: A Snowmass white paper. Journal of high energy astrophysics, 36, 55-110.
\bibitem{sources}
Spiering, C. (2012). Towards high-energy neutrino astronomy: a historical review. The European Physical Journal H, 37(3), 515-565.
\bibitem{crosssections}
Formaggio, J. A., \& Zeller, G. P. (2012). From eV to EeV: Neutrino cross sections across energy scales. Reviews of Modern Physics, 84(3), 1307-1341.
\bibitem{icecube}
Aartsen, M. G. et al. (2017). The IceCube Neutrino Observatory: instrumentation and online systems. Journal of Instrumentation, 12(03), P03012.
\bibitem{ic1}
Aartsen, M. G., et al. (2018). Differential limit on the extremely-high-energy cosmic neutrino flux... Physical Review D, 98(6), 062003.
\bibitem{ic2}
Detection of a particle shower at the Glashow resonance with IceCube. Nature, 2021, vol. 591, no 7849, p. 220-224.
\bibitem{ic3}
IceCube Collaboration, R. Abbasi et al., Science 378, 538 (2022), 2211.09972.
\bibitem{ic4}
Aartsen, M. G. et al. Science 361, eaat1378 (2018). arXiv preprint arXiv:1807.08816.
\bibitem{ic5}
R. Stein and other, Nature Astronomy 5, 510 (2021), 2005.05340
\bibitem{spb2}
Eser, J., Olinto, A. V., \& Wiencke, L. (2023). Overview and first results of EUSO-SPB2. arXiv preprint arXiv:2308.15693.
\bibitem{pueo}
Abarr, Q. et al. \& PUEO collaboration. (2021). The payload for ultrahigh energy observations (PUEO): a white paper. Journal of Instrumentation, 16(08), P08035.
\bibitem{pbr}
Battisti, M., Eser, J., Olinto, A., Osteria, G., \& JEM-EUSO Collaboration. (2024). POEMMA-Balloon with Radio. Nuclear Instruments and Methods in Physics Research Section A: Accelerators, Spectrometers, Detectors and Associated Equipment, 1069, 169819.
\bibitem{poemma}
Olinto, A. V., Krizmanic, J., ... \& Poemma Collaboration. (2021). The POEMMA observatory. Journal of Cosmology and Astroparticle Physics, 2021(06), 007.
\bibitem{flavours}
Bustamante, M., \& Ahlers, M. (2019). Inferring the flavor of high-energy astrophysical neutrinos at their sources. Physical Review Letters, 122(24), 241101.
\bibitem{oscillations}
Song, N. et al. (2021). Journal of Cosmology and Astroparticle Physics, 2021(04), 054.
\bibitem{astropy}
Robitaille, T. et al. (2013). Astropy: A community Python package for astronomy. Astronomy \& Astrophysics, 558, A33.
\bibitem{exitprob}
Garg, D. et al. (2023). Neutrino propagation in the Earth and emerging charged leptons with nuPyProp. Journal of cosmology and astroparticle physics, 2023(01), 041.
\bibitem{exitenergy}
Reno, M. H., Krizmanic, J. F., \& Venters, T. M. (2019). Cosmic tau neutrino detection via Cherenkov signals from air showers from Earth-emerging taus. Physical Review D, 100(6), 063010.
\bibitem{nupyprop}
 D. Garg et al., JCAP 2023, 041 (2023), 2209.15581.
\bibitem{nuleptonsim}
 Alvarez-Muñiz, J. et al. (2018). Comprehensive approach to tau-lepton production by high-energy tau neutrinos propagating through the Earth. Physical Review D, 97(2), 023021.
\bibitem{pythia}
Sjöstrand, T. (2020). The PYTHIA event generator. Computer Physics Communications, 246, 106910.
\bibitem{merra}
Gelaro, R., McCarty, W., Suárez, et al. (2017). The modern-era retrospective analysis for research and applications, version 2 (MERRA-2). Journal of climate, 30(14), 5419-5454.
\bibitem{zhaires}
Alvarez-Muniz, J., Carvalho Jr, W. R., \& Zas, E. (2012). Monte Carlo simulations of radio pulses in atmospheric showers using ZHAireS. Astroparticle Physics, 35(6), 325-341.
\bibitem{dask}
Rocklin, M. (2015, July). Dask: Parallel computation with blocked... In SciPy (pp. 126-132).

\bibitem{pip}
A. Reustle et al. NuSpaceSim pypi (pip) webpage, https://pypi.org/project/
 nuspacesim/ (2021).

\bibitem{heasarc}
$\nu$SpaceSim HEASARC web portal. https://heasarc.gsfc.nasa.gov/docs/nuSpaceSim/
\bibitem{github}
A. Reustle et al. NuSpaceSim github repository, https://github.com/NuSpaceSim/nuSpaceSim (2021).
\bibitem{conex}
Pierog, T., Alekseeva, et al. (2004). First results of fast one-dimensional hybrid simulation of EAS using CONEX. arXiv preprint astro-ph/0411260.
\bibitem{offline}
Argiro, S. et al. (2007). The offline software framework of the Pierre Auger Observatory. Nuclear Instruments and Methods in Physics Research Section A 580(3), 1485-1496.
\bibitem{eusooffline}
Abe, S. et al. (2024). EUSO-Offline: A comprehensive simulation and analysis framework. Journal of Instrumentation, 19(01), P01007.

\bibitem{taurunner}
Safa, I. et al. (2020). Journal of Cosmology and Astroparticle Physics, 2020(01), 012.
\bibitem{nupropearth}
Garcia, A. et al. (2020). Complete predictions for high-energy neutrino propagation in matter. Journal of Cosmology and Astroparticle Physics, 2020(09), 025.
\bibitem{chasm}
Buckland, I. J., \& Bergman, D. R. (2023). Universality of Cherenkov light in EAS. Astroparticle Physics, 150, 102832.
\bibitem{search}
Antares, I., Auger, P., Scientific, L. I. G. O., ..., \& Virgo. (2017). Search for high-energy neutrinos from binary neutron star merger GW170817 with ANTARES, IceCube, and the Pierre Auger Observatory.
\bibitem{gw17}
Hartley, W. (2017). Multi-messenger observations of a binary neutron star merger. The Astrophysical Journal Letters, 848(2), L12.








\end{thebibliography}
\end{document}